\documentclass{imsart}
\RequirePackage[OT1]{fontenc}
\RequirePackage{amsthm,amsmath}
\RequirePackage[numbers]{natbib}
\RequirePackage[colorlinks,citecolor=blue,urlcolor=blue]{hyperref}
\RequirePackage{hypernat}
\usepackage{amssymb,mathrsfs}
\usepackage{multirow}
\usepackage{comment}

\numberwithin{equation}{section}

\newtheorem{corollary}{Corollary}[section]
\newtheorem{lemma}{Lemma}[section]
\newtheorem{assumption}{Assumption}
\newtheorem{condition}{Condition}
\newtheorem{remark}{Remark}[section]
\newtheorem{example}{Example}
\newcommand{\Thm}[1]{\begin{theorem}\label{thm.#1}}
\newcommand{\Cor}[1]{\begin{corollary}\label{cor.#1}}
\newcommand{\Prop}[1]{\begin{proposition}\label{prop.#1}}
\newcommand{\Lm}[1]{\begin{lemma}\label{lm.#1}}
\newcommand{\Ass}[1]{\begin{assumption}\label{ass.#1}}
\newcommand{\Ex}[1]{\begin{example}\label{ex.#1}\rm}
\newcommand{\Rem}[1]{\begin{remark}\label{rem.#1}\rm}
\newcommand{\thmref}[1]{Theorem~\ref{thm.#1}}

\newcommand{\Cond}[1]{\begin{condition}\label{cond.#1}\rm}

\def\qed{\hfill \mbox{\raggedright \rule{0.1in}{0.1in}}}

\def\Var{\mathop{\rm Var}\nolimits}%
\def\R{\mathbb R}
\def\bea{$$\begin{aligned}}
\def\eea{\end{aligned}$$}

\def\implies{{\Longrightarrow}}

\def\und{\quad \mbox{and} \quad}

\def\1{\mathbf 1}
\def\ave{n^{-1}\sum_{j=1}^n}

\def\Ave{\frac{1}{n}\sum_{j=1}^n}

\def\lam{\lambda} 

\def\maxj{\max_{1\leq j \leq n}}

\renewcommand{\Ref}[1]{(\ref{#1})}
\newcommand{\bel}[1]{\begin{equation}\label{#1}}
\newcommand{\tel}[1]{\begin{equation}\tag{#1}}
\newcommand{\ee}{\end{equation}}
\newcommand{\bed}[1]{\begin{eqnarray}\label{#1}}
\newcommand{\eed}{\end{eqnarray}}

\newcommand{\half}{\frac{1}{2}}

\newtheorem{theorem}{Theorem}[section]

\newcommand{\beds}{\begin{eqnarray*}}
\newcommand{\eeds}{\end{eqnarray*}}

\def\Z{{\mathcal Z}}

\def\V{\mathscr{V}} 

\def\S{\mathbb S}

\def\Normal{\mathscr{N}}

\def\bF{\mathbb F}

\def\Z{\mathscr{Z}}

\def\prob{\mathscr{P}_n}
\newcommand{\sel}[1]{\sup\Big\{ \prod_{j=1}^n n\pi_j: \bpi \in \prob,\  #1 \Big\}}
\newcommand{\isel}[1]{\sup\Big\{ \prod_{i=1}^n n\pi_i: \bpi \in \prob,\  #1 \Big\}}

\newcommand{\selN}[1]{\sup\Big\{ \prod_{j=1}^{N_n} N_n\pi_j: \bpi \in \mathscr{P}_{N_n},\  #1 \Big\}}

\def\H{\mathbb H}

\def\vep{{\varepsilon}}

\def\bdm#1{\mbox{\boldmath$#1$\unboldmath}}

\def\btheta{\mbox{\boldmath$\theta$\unboldmath}}

\def\bpi{\bdm{\pi}}

\startlocaldefs
\numberwithin{equation}{section}

\numberwithin{equation}{section}
\theoremstyle{plain}
\def\Var{\mathop{\rm Var}\nolimits}%
\def\R{\mathbb R}
\def\x{\times}

\def\prob{\mathscr{P}_{n}}
\def\und{\quad \mbox{and} \quad}

\def\1{\mathbf 1}
\def\Ave{\frac{1}{n}\sum_{j=1}^n}
\def\ave{n^{-1}\sum_{j=1}^n}

\def\iAve{\frac{1}{n}\sum_{i=1}^n}

\def\maxj{\max_{1\leq j \leq n}}

\def\Z{{\mathbf Z}}

\def\V{\mathscr{V}}

\def\S{\mathbb S}

\def\P{\mathscr{P}}

\def\U{\mathbb{U}_n}

\newcommand{\ssel}[1]{\sup\Big\{ \prod_{j=1}^n n\pi_j: &
(\pi_1,\dots,\pi_n)\in \Pi_n, \\ & #1 \Big\}}

\def\L{\mathrm{L}}
\def\b{\mathbf{b}}
\def\u{\mathbf{u}}
\def\jsum{\sum_{j=1}^n}
\def\htet{\hbtheta}
\def\ttet{\tbtheta}
 
\def\seq#1{\left\{#1\right\}}

\def\lam{\lambda}

\def\seq#1{\left\{#1\right\}}
\def\V{\mathscr {V}}

\def\pare#1{\left(#1\right)}

\def\Normal{\mathscr{N}}

\def\bA{{\mathbb{A}}}

\def\bF{{\mathbb{F}}}
\def\bG{{\mathbb{G}}}
\def\bH{{\mathbb{H}}}
\def\bI{{\mathbb{I}}}

\def\C{{\mathbf{C}}}

\def\H{{\mathbf{H}}}
\def\I{{\mathbf{I}}}
\def\J{{\mathbf{J}}}
\def\K{{\mathbf{K}}}
\def\L{{\mathbf{L}}}
\def\M{{\mathbf{M}}}

\def\P{{\mathbf{P}}}

\def\R{{\mathbf{R}}}

\def\U{{\mathbf{U}}}
\def\V{{\mathbf{V}}}
\def\W{{\mathbf{W}}}
\def\X{{\mathbf{X}}}

\def\Z{{\mathbf{Z}}}

\def\cR{{\mathcal{R}}}

\def\cT{{\mathcal{T}}}

\def\cZ{{\mathcal{Z}}}

\def\a{{\mathbf{a}}}
\def\b{{\mathbf{b}}}
\def\c{{\mathbf{c}}}

\def\f{{\mathbf{f}}}

\def\m{{\mathbf{m}}}

\def\r{{\mathbf{r}}}
\def\s{{\mathbf{s}}}
\def\t{{\mathbf{t}}}
\def\u{{\mathbf{u}}}
\def\v{{\mathbf{v}}}

\def\x{{\mathbf{x}}}
\def\y{{\mathbf{y}}}

\def\seq#1{\left\{#1\right\}}
\def\pare#1{\left(#1\right)}

\def\bdm#1{\mbox{\boldmath$#1$\unboldmath}}
\def\btheta{\mbox{\boldmath$\theta$\unboldmath}}
\def\hbtheta{\bdm{\hat\btheta}}
\def\tbtheta{\bdm{\tilde\btheta}}

\def\bpi{\bdm{\pi}}
\def\bzeta{\bdm{\zeta}}
\def\hbzeta{\bdm{\hat\zeta}}
\def\tbzeta{\bdm{\tilde\zeta}}

\def\bpsi{\bdm{\psi}}

\def\bphi{\bdm{\phi}}
\def\bvarphi{\bdm{\varphi}}

\def\qed{$\hfill\square$}
\def\ob{s}
\def\bob{\mathbf{s}}
\def\odd{\mathrm{odd}}

\begin{document}
\renewcommand{\Ref}[1]{(\ref{#1})}

\begin{frontmatter}
\title{Easy Maximum Empirical Likelihood Estimation of Linear Functionals 
Of A Probability Measure With Infinitely Many Constraints}
\runtitle{An EL-approach of Estimation of Linear Functionals}

\begin{aug}
\author{\fnms{Shan} \snm{Wang} \thanksref{th1}
	\ead[label=e1]{swang151@usfca.edu}}
\and
\author{\fnms{Hanxiang} \snm{Peng}
	 \ead[label=e2]{hanxpeng@iu.edu}}
\thankstext{th1}{Corresponding author}
\runauthor{S. Wang and H. Peng}

\address{ University of San Francisco \\ 
         Department of Mathematics and Statistics\\
         San Francisco, CA 94117, USA \\
         swang151@usfca.edu
         \printead{e1}
}

\address{ Indiana University Purdue University Indianapolis \\
         Department of Mathematical Sciences\\
         Indianapolis, IN 46202-3267, USA \\
         hanxpeng@iupui.edu
         \printead{e2}
}

\end{aug}

\begin{abstract}
In this article, 
we construct semiparametrically efficient estimators of linear functionals 
of a probability measure in the presence of side information using an easy empirical likelihood
approach. We use estimated constraint functions and allow the number of constraints to grow with the sample size. Considered are three cases of information which can be
characterized by infinitely many constraints: (1) the marginal distributions 
are known, (2) the marginals are unknown but identical, 
and (3) distributional symmetry. 
An improved spatial depth function is defined 
and its asymptotic properties are studied. Simulation results on efficiency gain are reported. 
\end{abstract}



\end{frontmatter}

\section{Introduction}
\label{intro}
Suppose that $Z_1, \dots, Z_n$ are independent and identically distributed (i.i.d.) random variables with a common distribution 
$Q$ taking values in a measurable space $\cZ$. In this article, we are interested 
in efficient estimation of the linear functional $\btheta=\int \bpsi\,dQ$ of $Q$ 
for some square-integrable function $\bpsi$ from $\cZ$ to $\cR^r$ when side information is available 
through a vector function (constraint) $\u$  which satisfies     
\begin{itemize}
\item[(C)] $\u$ is measurable from $\cZ$ to $\cR^m$ such that $\int \u\,dQ=0$ and
the variance-covariance matrix $\int\u\u^\top\,dQ$ is nonsingular.
\end{itemize}
 
  The commonly used sample mean $\bar\bpsi=\Ave \bpsi(Z_j)$ of $\btheta=E(\bpsi(Z))$
does not use the  information,  
and is not efficient in the sense of least dispersed regular estimators, 
see e.g. Bickel, Klaassen, Ritov and Wellner (1993). 
  Based on the criterion of maximum empirical likelihood, an improved estimator 
which utilizes  the  information is 
\bel{tele}
\tbtheta=\Ave \frac{\bpsi(Z_j)}{1+\u(Z_j)^\top \tbzeta}, 
\ee 
where $\tbzeta$ is the solution to the equation
\bel{tele2}
\sum_{j=1}^n \frac{\u(Z_j)}{1+ \u(Z_j)^\top \bzeta}=0. 
\ee 
We shall refer to $\tbtheta$ as the \emph{EL-weighted} estimator.
   
   There is an extensive amount of literature on the empirical likelihood testing of 
hypothesis, see e.g. Owen (1988, 2001). 
Soon it was used to construct point estimators. 
Qin and Lawless (1994) studied maximum empirical likelihood estimators (MELE) and showed
in Corollary 2 that MELE are fully efficient.
As a special case of MELE, estimators of the preceding easy form were studied 
in Zhang (1995,  1997) in M-estimation and quantile processes in the presence of 
auxiliary information (side information).   
  For a \emph{fixed} number $m$ of \emph{known} constraint functions, the asymptotic normality (ASN) and efficiency of MELE were established. 

    Hjort, McKeague and Van Keilegom (2009) extended the scope of the empirical likelihood testing hypothesis, and developed a general theory for constraints with nuisance parameters and considered the case with infinitely many constraints.     Peng and Schick (2013) generalized the empirical likelihood testing to
allow for the number of constraints to grow with the sample size and for the constraints 
to use estimated criteria functions. 
   Peng and Tan (2018) expanded the results of the latter to U-statistics based 
general estimating equations with side information. 

    Parente and Smith (2011)  studied generalized empirical likelihood estimators 
for irregular constraints. 
Peng and Schick (2018) presented a theory of maximum empirical likelihood
estimation and empirical likelihood ratio testing with irregular and 
estimated constraint functions.
   Wang and Peng (2022) used the easy EL-weighted approach to construct 
improved estimators of linear functionals of a probability measure 
when side information is available. 
Motivated by nuisance parameters common in semiparametric
models and the infinite dimension of such models, they studied the use of estimated functions for growing number of constraints with the sample size. 
They applied the results to improve estimation efficiency 
in the structural equation models. 

We shall rely the results of Wang and Peng (2022) to construct efficient estimators
of linear functionals of a probability measure for a few cases of side information
which is determined by infinitely many constraints. 
Bickel, Ritov and Wellner  (1991) characterized efficient estimation of $E(h(X; Y ))$
for known $h$ when the marginal distributions of $X$ and of $Y$ are \emph{known},
and construct an efficient estimator based on the criterion of minimum chisquare-type objective
function.
Peng and Schick (2005) calculated the information lower bound when the marginal distributions
are unknown but identical, and constructed an efficient estimator based on the criterion
of least squares objective. 
Peng and Schick (2018) constructed empirical likelihood tests of stochastic independence
and distributional symmetry. 
Each of independence, symmetry, known or equal marginal distributions is equivalent to 
infinitely many equations (constraints), and can be used to improve estimation efficiency. 
Here we construct the EL-weighted estimators and demonstrate the semiparametric efficiency. 
Note the simple analytic form of our estimators, and the property of easy   
incorporation of side information to improve efficiency.  


The efficiency criteria used are that of a least dispersed regular estimator or that of a locally asymptotic minimax estimator, and are based on the convolution theorems and
on the lower bounds of the local asymptotic risk in LAN and LAMN families,
see the monograph by Bickel, et al. (1993) among others. 

  In what follows, we will summarize some results from Wang and Peng (2022) for 
the convenience of our use. Meanwhile, we provide the proof of the semiparametric effiency. 
  In many semiparametric models, the constraint vector function 
$\u=(u_1, ..., u_m)^\top$ is usually unknown and must be estimated by some 
measurable function $\hat \u=(\hat u_1, ..., \hat u_m)^\top$. Using it, we now work with the EL-weights,   
\bel{ho1}
\hat\pi_j=\frac{1}{n}\frac{1}{1+\hat\u(Z_j)^\top \hbzeta}, \quad j=1,\dots,n,
\ee
where $\hbzeta$ solves Eqt \Ref{tele2} with $\u=\hat \u$.
   A natural  estimate $\htet$ of $\btheta$ now is 
\bel{hele}
\htet=\jsum \hat\pi_j\bpsi(Z_j)=\Ave \frac{\bpsi(Z_j)}{1+\hat \u(Z_j)^\top \hbzeta}.
\end{equation}

 We now allow the number of constraints to depend on the sample size $n$, 
 $m=m_n$,  and tend to infinity slowly with  $n$. To stress the dependence, 
write 
$$\u_n=(u_1, \dots, u_{m_n})^\top, \quad
\hat \u_n=(\hat u_1, \dots, \hat u_{m_n})^\top, 
$$ 
and $\ttet_n=\ttet$, $\htet_n=\htet$ for 
the corresponding estimators of $\btheta$, that is,
\bel{telen}
\ttet_n=\Ave \frac{\bpsi(Z_j)}{1+\u_n(Z_j)^\top \tilde\bzeta_n} \und
\htet_n=\Ave \frac{\bpsi(Z_j)}{1+\hat\u_n(Z_j)^\top \hat\bzeta_n},
\end{equation}   
where $\tilde\bzeta_n$ and $\hat\bzeta_n$ solve  
Eqt \Ref{tele2} with $\u=\tilde\u_n$ and $\u=\hat\u_n$,  respectively,.

  The ASN of $\ttet_n$ and $\htet_n$ are, respectively, given in Theorems 3 and 4 of 
Wang and Peng (2022), 
and we now prove the semiparametric efficiency of $\ttet_n$ and quote Theorem 4 in the Appendix
for convenience of our use. 
For $\a \in \cR^m$, write $\|\a\|$ the euclidean norm. 
For $\a, \b \in \cR^m$, write $\a\otimes \b$ the kronecker product. 
Let $L_2^m(Q)=\seq{\f=(f_1, \dots, f_m)^\top: \int \|\f\|^2\,dQ<\infty}$, and let
$L_{2,0}^m(Q)=\seq{\f \in L_2^m(Q): \int \f\,dQ=0}$.
For $\f \in L_2^m(Q)$, write $\bar \f=\ave \f(Z_j)$ the sample average of $\f(Z_1), \dots, \f(Z_n)$, 
and $[\f]$ the closed linear span of the components $f_1, \dots, f_r$ in $L_2(Q)$. 
Let $Z$ be an i.i.d. copy of $Z_1$.  
%
%
Denote by $[\u_\infty]$ the closed
linear span of $\u_\infty=(u_1, u_2, \dots)$ in $L_{2,0}(Q)$. Set 
\[
\W_n=\Var(\u_n(Z)), \quad
\bar \W_n=\Ave (\u_n\u_n^\top)(Z_j), \quad 
\hat \W_n=\Ave (\hat\u_n\hat\u_n^\top)(Z_j).
\] 
Following Peng and Schick (2013), a sequence $\W_n$ of $m_n\times m_n$ dispersion matrices is said to be \emph{regular} if
$$
0 < \inf_n \inf_{\|\u\|=1} \u^{\top}\W_n \u \leq \sup_n \sup_{\|\u\|=1} \u^{\top}\W_n \u <\infty.
$$

\Thm{3} 
Suppose that $\u_{n}$ satisfies (C) for each $m=m_n$  
such that
\bel{3a}
\maxj \|\u_n(Z_j)\|=o_p(m_n^{-3/2}n^{1/2}), 
\ee 
the sequence of $m_n \times m_n$ dispersion matrices $\W_n$ is regular and satisfies
\bel{3b}
|\bar \W_n-\W_n|_o=o_p(m_n^{-1}), 
\ee
\bel{3c}
\Ave\pare{\bpsi(Z_j)\otimes \u_n(Z_j)-E\big(\bpsi(Z_j)\otimes \u_n(Z_j)\big)}=o_p(m_n^{-1/2}).
\ee
Then  $\ttet_n$ is semiparametrically efficient as $m_n\to \infty$. Moreover,  
$$
\sqrt{n}(\ttet_n-\btheta) \implies \Normal(0, \varSigma_0),
$$
where $\varSigma_0=\Var(\bpsi(Z))-\Var(\bvarphi_0(Z))$ with 
$\bvarphi_0=\Pi(\bpsi|[\u_\infty])$. 
\end{theorem} 
{\sc Proof}. 
We only need to show the efficiency. It suffices to prove that the orthonormal complement
$\cT=[\u_\infty]^\perp$ in $L_{2,0}(Q)$ is the tangent space. 
To this end, let  $Q_t: |t|\leq t_0$ with $Q_0=Q$ be a regular parametric submodel 
with the score function $a$. By (C),
$$
\int u\,dQ_t=0, \quad u\in[\u_\infty].
$$
Differentiating both sides of the equality with respect to $t$ at $t=0$ yields
$$
\int ua\,dQ=0, \quad u\in[\u_\infty].
$$ 
This shows $a\in\cT$. 
For any bounded $a\in\cT$, consider  
$q_t=dQ_t/dQ=1+at, |t|\leq t_0$ for sufficient small $t_0$. It is clear that  
$q_t$ is a density and the submodel with the density has the score function 
$a$ which satisfies $\int ua\,dQ=0$. Since bounded functions in $\cT$ are dense, 
it follows that the above conclusion holds for any $a \in \cT$. This shows 
$\cT$ is the tangent space. 
\qed

The article is organized as follows. In Section \ref{spd}, the EL-weighted 
spatial depth function is constructed, and its ASN and efficiency are 
established in the presence of distributional symmetry.
 The ASN and efficiency of the EL-weighted estimators of linear functionals
are proved when the marginal distribution functions are known
in Section \ref{KM}, and when the marginal distributions are unknown
but equal in Section \ref{EQM}. The simulation results are reported in 
Section \ref{sim}. Section \ref{app} contains Theorem 4 of Wang and Peng (2022).

\def\M{\mathbf{m}}
\section{The EL-weighted spatial median}\label{spd}
 In this section, we introduce the EL-weighted spatial depth function, 
exhibit efficiency and give the asymptotic normality.   

  The statistical depth functions provide a center-outward ordering of a point in $\cR^p$ with respect to a distribution. High depth values correspond to centrality while low values to ``outlyingness''. 
Depth functions possess robustness property, and can be used to define multivariate medians, 
which are robust location estimators. 
Common depth functions include the Tukey depth (halfspace depth), 
the simplicial depth, the projection depth, and the spatial depth.  
 Here we shall use the easy EL-approach to constructing improved depths, and  
illustrate it with the spatial depth. 
The (population) spatial depth function $D(\x)$ with respect to a distribution $F$ is defined as 
$$
D(\x)=1-\|E\big( \S(\x-\X)\big)\|, \quad \x \in \cR^p,
$$
where $\S(\x)=\x/\|\x\|$ if $\x\neq 0$ ($\S(0)=0$) is the spatial sign function
and $\X$ has the distribution function (DF) $F(\x)$, denoted by $\X\sim F(\x)$. 
The depth function $D(\x)$ can be estimated by the 
sample depth function given by 
$$
D_n(\x)=1-\Big\|\iAve \S_\x(\X_i)\Big\|.
$$
where $\S_\x(\t)=\S(\t-\x)$.
The sample spatial median $\M_n$ is defined as the value which maximizes the depth function, that is, 
$$
\M_n=\arg \max_{\x\in \cR^p} D_n(\x)=\arg \min_{\x\in \cR^p} \Big\|\iAve \S_\x(\X_i)\Big\|. 
$$

 Suppose that there is available additional information that can be expressed 
by a constraint function $\u$. 
While the sample depth  $D_n(\x)$ does not utilize the information,
the \emph{EL-weighted depth function} $\widetilde D_n(\x)$ makes use of it 
and is defined by  
\bel{dpf}
\widetilde D_n(\x)
=1-\Big\|\iAve \frac{\S_\x(\X_i)}{1+\u(\X_i)^\top \tbzeta}\Big\|, \quad \x \in \cR^p, 
\ee
where $\tbzeta$ is the solution to the equation
\bel{31}
\sum_{j=1}^n\frac{\u(\X_j)}{1+\u(\X_j)^\top\bzeta}=0.
\ee

The EL-weighted spatial median $\widetilde \M$ is defined 
as the value which maximizes the EL-weighted depth function, that is, 
\bel{elsm}
\widetilde\M=\arg \max_{\x\in \cR^p} \widetilde D_n(\x)
=\arg \min_{\x\in \cR^p} \Big\|\iAve \frac{\S(\x-\X_i)}{1+\u(\X_i)^\top \tbzeta}\Big\|. 
\ee
The EL-weighted estimator of $\btheta(\x)=E(\S_\x(\X))$ is  given by 
 \bel{df1}
 \tbtheta(\x)=\iAve \frac{\S_\x(\X_i)}{1+\u(\X_i)^\top \tbzeta}, \quad \x\in\R^p. 
 \ee

\Rem{rob} The sample spatial $D_n(\x)$ is robust with the breakdown point $1/2$.
The EL-weighted $\widetilde D_n(\x)$ improves efficiency but reduces robustness resulted from the
zero value of the EL-weights. One can robustify $\widetilde D_n(\x)$ by truncating the EL-weights
from below by a fixed constant. Truncation is commonly used in the inverse probability weighing
method. Obviously, truncation leads to certain loss of efficiency. 
\end{remark}

   {\bf Known marginal medians}.  
In our simulation study, we looked at the side information that the bivariate random vector
$\X=(X_1, X_2)^\top$ has \emph{known} marginal medians $m_{10}$ and $m_{20}$. 
That is, the componentwise median $(m_{10},m_{20})^\top$ is known.   
In this case, $\u(x_1, x_2)=(\1[x_1 \leq m_{10}]-1/2, \1[x_2\leq m_{20}]-1/2)^\top$. 
 We are motivated as follows.  
 It is well known that the spatial median is a better location estimator than the componentwise median because the former takes into account the correlation of the components while the latter ignores it, see Chen, Dang, Peng and Bart (2009). We are interested in
how much information is lost when the componentwise median is used
by looking at how much efficiency of the EL-weighted spatial median $\tilde\M$ 
(when the marginal medians are known)
gains over the sample spatial median (when the marginal medians are unknown).

 {\bf Growing number of constraints}. 
Suppose that there exists some constant vector $\c$ such that 
$T=\c^{\top}\X$ is \emph{symmetric} about some known value $\tau_0$.
Let $\varepsilon_j=\c^{\top}\X_{j}-\tau_0, j=1, \ldots, n$. 
Then $\vep_j$'s are i.i.d. random variables which are symmetric about zero. 
Let $\vep$ be an i.i.d. copy of $\vep_j$'s,  and let $F$ be the distribution function of $\vep$. 
Let $L_{2,0}(F, \odd)$ be the subspace of $L_{2,0}(F)$ consisting of the odd functions. Symmetry of $\varepsilon$ about $0$ implies
$$
E(a(\varepsilon))=0, \quad a\in L_{2,0}(F, \odd).
$$
Let $\ob_{k}(t)=\sin(k\pi t), t\in [-1,1], k=1,2, ...$ be the orthonormal 
trigonometric basis.  
Define $G(t)=2F(t)-1, t \in \cR$. Then $G(t)$ is an odd function in $L_{2,0}(F, \odd)$, 
and $\ob_{k}(G(t)), k=1, 2, ...$ form a basis of the space.

 In this case, the constraints are $\u_n(\X_j)=(\ob_1(G(\varepsilon_j)), ..., \ob_{m_n}(G(\varepsilon_j)))^{\top}$, 
where we allow $m_n$ to grow to infinity slowly with $n$. 
The EL-weighted depth function is calculated by \Ref{dpf} with $\u=\u_n$ and 
$\tbzeta=\tbzeta_n$ which solves Eqt \Ref{31} with $\u=\u_n$.
The EL-weighted estimator of $\btheta(\x)=E(\S_\x(\X))$ then is 
 \bel{df1}
 \tbtheta_n(\x)=\iAve \frac{\S_\x(\X_i)}{1+\u(\X_i)^\top \tbzeta_n}, \quad \x\in\R^p. 
 \ee

\Thm{df} Suppose that $F$ is continuous. 
Then for arbitrary but fixed $\x\in\R^p$, as $m_n \to \infty$  such that $m_n^4/n\to 0$, $\ttet_n(\x)$ in \Ref{df1} satisfies 
$$
\ttet_n(\x)=\bar \S_\x-\bar\bvarphi_{\x0}+o_p(n^{-1/2}),
$$ 
where $\bvarphi_{\x0}=\Pi(\S_\x(\X)|L_{2,0}(F,\odd))$ is the projection of $\S_\x(\X)$ onto $L_{2,0}(F,\odd)$. 
As a consequence, if $\varSigma_{0}(\x)=\Var(\S_\x(\X))-\Var(\bvarphi_{\x0}(\X))$ is nonsingular,   
$$
\sqrt{n}(\ttet_n(\x)-\btheta(\x)) \implies \Normal(0, \varSigma_0(\x)).
$$

\end{theorem}  

{\sc Proof of \thmref{df}}. 
 We shall apply \thmref{3} to prove the result. 
Since $\W_n=E((\u\u^\top)(\X))=\I_{m_n}$ is the identity matrix, it follows that (C) holds 
and $\W_n$ is regular.  As $\| \u_n(\X_j)\| \leq m_n^{1/2}$ for each $j$ and $m_n^4/n=o(1)$,  
\Ref{3a} is satisfied, while \Ref{3b} holds in view of the inequalities   
$$
nE(|\bar \W_n-\W_n|_o^2)
\leq E(\|\u_n(\X)\|^4)
\leq m_n^2. 
$$
 Let $\K_n$ be the left hand side of \Ref{3c}. Then \Ref{3c} follows from
$$
nE(\|\K_n\|^2)\leq  E(\|\S_\x(\X)\otimes\u_n(\X)\|^2)
\leq {m_n}E(\|\S_\x(\X)\|^2)={m_n}.
$$
We now apply \thmref{3} to complete the proof. 
\qed
\medskip 

{\bf Efficiency gain and ASN for $\tilde \M$}. 
By the properties of empirical likelihood, one concludes that $\widetilde D_n(\x)$ is a valid depth function at least for large $n$ as  all $1+\u(\X_i)^\top \tbzeta >0$.  
Fix $\x \in \cR^p$,  let $\P_{\x}$ be the projection of $\S_\x(\X)$ onto the closed linear span $[\u_\infty]=L_{2,0}(F,odd)$.
Then $\varSigma_0(\x)=\Var(\S_\x(\X))-\Var(\P_{\x}(\X))$. Clearly,
\bel{spd1}
\Var(\P_{\x}(\X))=E\big(\S_\x(\X)v\otimes \u(\X)^\top\big)\W^{-1}E\big(\S_\x(\X)\otimes \u(\X)\big).
\ee
Let $\S_2(\x)=\S\big(E(\S_\x(\X))\big)$.
If $\W_0(\x):=\S_2(\x)\V_0(\x)\S_2(\x)^\top$ is nonsingular, then by \thmref{df} 
for fixed $\x\in\R$, 
\[
\sqrt{n}(\widetilde D_n(\x)-D(\x))\implies \Normal(0, \W_0(\x)).
\]
 Note that the sample depth $D_n(\x)$ satisfies
\[
\sqrt{n}(D_n(\x)-D(\x))\implies \Normal(0, \W(\x)),
\]
where $\W(\x)=\S_2(\x)\Var(\S_\x(\X))\S_2(\x)^\top$.
  Thus the reduction of the asymptotic variance-covariance of the EL-weighted depth
$\widetilde D_n(\x)$ is  
$$
\S_2(\x)\Var(\P_{\x}(\X))\S_2(\x)^\top.
$$ 

 We now use the Delta method to drive the ASN of the EL-weighted 
spatial median $\tilde\M$.
To this end, we need some results from Chaudhuri (1992) in the case of $m=1$ for which 
the spatial median corresponds to his multivariate Hodges-Lehmann type location estimate. 
The following is his Assumption 3.1. 
\begin{itemize}
\item[(PC)]   
 $\X_{1}$, ${\dots}$, $\X_{n}$ are i.i.d random vectors in $\cR ^{d}$ with an absolutely continuous (with respect to the Lebesgue measure) distribution having a density $f$ that is bounded on every bounded subset of $\cR ^{d}$.  
\end{itemize} 
Assume (PC) and $d\ge 2$. 
 Let $\H(\x)=\|\x\|^{-1}(\I_d-\x\x^\top/\|\x\|^2)$ if $\x\neq 0$ and
$\H(0)=0$.  Note that $\S(\x)$ and $\H(\x)$ are the first and second order 
partial derivatives of $\|\x\|$.     
Under (PC), the underlying distribution is absolutely continuous with respect to the Lebesgue measure
on $\cR^p (d\ge 2)$, hence the (population) spatial median $\M_0$ uniquely exists and satisfies the equation $E(\S(\M_0-\X))=0$. The spatial median $\M_n$ satisfies  
$$
\sum_{i=1}^n \S(\M_n-\X_i)=0.
$$
Let $\J=E\big((\S\S^\top)(\m_0-\X)\big)$ and $\K=E\big(\H(\M_0-\X)\big)$.
Chaudhuri (1992) showed in his Theorem 3.3 and its corollary that if (PC) holds then the matrices
$\J$ and $\K$ are positive definite and $\M_n$ satisfies
$$
\sqrt{n}(\M_n-\M_0)\implies \Normal(0,\, \K^{-1}\J\K^{-\top}).
$$ 
Note that  the EL-weighted spatial median $\widetilde \M_n$ satisfies the equation,  
$$
\sum_{i=1}^n \frac{\S(\M-\X_i)}{1+\u(\X_i)^\top\tbzeta}=0.
$$ 
Using the Delta method, we derive, with $\V_0(\M_0)=\J-\Var(\P_{\M_0}(\X))$,
$$
\sqrt{n}(\widetilde\M_n-\M_0)\implies \Normal(0,\, \K^{-1}\V_0(\M_0)\K^{-\top}), 
$$ 
where $\Var(\P_{\M_0}(\X))$ is calculated by \Ref{spd1}.

 {\bf Growing number of estimated constraints}. For unknown $F(x)$, we estimate it by the symmetrized empirical distribution function,  
$$
\bF(x)=\Ave \frac{\1[\varepsilon_j\leq x]+\1[-\varepsilon_j\leq x]}{2}, \quad x \in \cR.
$$
Let $\bG(x)=2\bF(x)-1$. We thus obtain computable functions $\ob_j(\bG(x))$. 
Write $\u_n$ for $\u$, and  
estimate it by $\hat \u_n(x)=(\ob_1(\bG(x)), ..., \ob_{m_n}(\bG(x)))^{\top}$. 
 The EL-weighted estimator of $\btheta(\x)=E(\S_\x(X))$ is now given by  
\bel{df2}
\htet_n(\x)=\iAve \frac{\S_\x(\X_i)}{1+\hat \u_n(\X_i)^\top \hbzeta_n}, \quad \x \in \cR^p, 
\ee
where $\hbzeta_n$ solves Eqt \Ref{31} with $\u=\hat\u_n$. We have 

\Thm{dff} Suppose that $F$ is continuous.
Then $\htet_n$ defined in \Ref{df2} satisfies the conclusions of \thmref{df} 
as $m_n \to \infty$  such that $m_n^6/n\to 0$. 
\end{theorem} 

{\sc Proof of \thmref{dff}}.  We shall use \thmref{4} of Wang and Peng (2022) for the proof. 
First,  (C) is satisfied with $\W_n$ regular as $\W_n=I_{m_n}$. Next,  
\Ref{4a} follows from $\|\hat\u_n(Z_j)\|^2\leq m_n$ and $m_n^4/n=o(1)$.
Let
$$
\hat \W_n=\Ave (\hat \u_n\hat \u_n^\top)(\X_j), \quad
\bar \W_n=\Ave  (\u_n\u_n^\top)(\X_j)^\top. 
$$
Then $\bar \W_n-\W_n=o_p(m_n^{-1})$ follows from $m_n^4/n=o(1)$
and 
$$
nE(|\bar \W_n-\W_n|_o^2)
\leq E(\|\u_n(\Z_1)\|^4)
\leq m_n^2.
$$
Let $D_n=\ave\|\hat\u_n(\X_j)-\u_n(\X_j)\|^2$. It is easy to see 
$$
|\hat \W_n-\bar \W_n|_o
\leq D_n+2|\bar \W_n|_o^{1/2}D_n^{1/2}.
$$
Thus \Ref{4b} follows from 
\( 
D_n=o_p(m_n^{-2})
\)
to be shown next. To this end, let $\bob_n=(\ob_1, ..., \ob_{m_n})^\top$. Then  $\|\bob_n(t)\|\leq m_n^{1/2}$. 
One verifies $\|\bpsi_n'(t)\|\leq am_n^{3/2}$ for some constant $a$. Therefore,
\(D_n=o_p(m_n^{-2})\) follows from $D_n=O_p(m^3_n/n)$ and 
$m_n^5/n=o(1)$, in view of 
$$
\Ave \|\bob_n(\bG(\X_j))-\bob_n(G(\X_j))\|^2\leq a m_n^3\sup_{t\in\cR}|\bG(t)-G(t)|^2
=O_p(m^3_n/n).
$$  
   Denoting $\bpsi(\y)=\S_\x(\y)$, we break 
\[
\Ave \Big(\bpsi(\X_j)\otimes \hat \u_n(\X_j)-E\big(\bpsi(\X_j)\otimes \u_n(\X_j))\Big)
=\J_n+\K_n,  \quad \textrm{where} 
\]
$$
\begin{aligned}
& \J_n=\Ave \bpsi(\X_j)\otimes \big(\hat \u_n(\X_j)- \u_n(\X_j)\big),\\
& \K_n=\Ave \Big(\bpsi(\X_j)\otimes \u_n(\X_j)- E\big(\bpsi(\X_j)\otimes\u_n(\X_j)\big)\Big).
\end{aligned}
$$
By Cauchy inequality, 
$$
\begin{aligned}
 E(\|\J_n\|^2)
&\leq E(\|\bpsi(\X_1)\|^2) \Ave E(\|\hat \u_n(\X_j)-\u_n(\X_j)\|^2)\\
&=E(D_n)
=O(m_n^3/n)=o(m_n^{-1}),
\end{aligned}
$$
as $m_n^4/n=o(1)$. 
We now bound the variance by the second moment
to get
$$
E(\|\K_n\|^2)\leq  \frac{1}{n}E(\|\bpsi(\X_1)\otimes\u_n(\X_1)\|^2)
\leq 4\frac{m_n}{n}=o(m_n^{-1})
$$
as $m_n^2/n=o(1)$. Taken together we prove \Ref{4c} -- \Ref{4d}. 

 We now show that \Ref{4e} holds with $\v_n=\u_n$. To this end, 
using Taylor expansion we write
$$
\Ave (\hat\u_n(\X_j)-\u_n(\X_j))
=\L_n+\M_n, \quad \textrm{where}
$$
$$
\L_n=\Ave \bpsi_n'(G(\varepsilon_j))\big(\bG(\varepsilon_j)-G(\varepsilon_j)\big),\,
\M_n=\Ave \bpsi_n''(G_{nj}^*)\big(\bG(\varepsilon_j)-G(\varepsilon_j)\big)^2,
$$
where $G_{nj}^*$ lies in between $\bG(\varepsilon_j)$ and $G(\varepsilon_j)$. 
It thus follows  
$$
\begin{aligned}
E\big(\|\L_n\|^2\big)
&\leq \frac{1}{n} E\Big(\|\bpsi_n'(G(\varepsilon_1))\|^2\big(\bG(\varepsilon_j)-G(\varepsilon_j)\big)^2\Big)\\
&\leq a \frac{m_n^3}{n} \sup_{t \in \cR}\big(\bG(t)-G(t)\big)^2\\
&=O_p(m_n^3/n^2)=o_p((m_nn)^{-1})
\end{aligned}
$$ 
as $m_n^4/n=o(1)$.  This shows $\L_n=o_p((m_nn)^{-1/2})$. 
 One has $\|\bpsi''(t)\|=O_p(m_n^{5/2})$. Using this, we get 
$$
\|\M_n\|\leq O(m_n^{5/2})\sup_{t\in\cR}|\bG(t)-G(t)|^2
=O_p(m_n^{5/2}/n)=o_p((m_nn)^{-1/2})
$$
as $m_n^6/n=o(1)$. This yields $\M_n=o_p((m_nn)^{-1/2})$.
Taken together the desired \Ref{4e} follows. 
We now apply \thmref{4} to finish the proof. 
\qed


\def\M{\mathbf{M}}

\section{Efficient estimation of linear functionals with known marginals}\label{KM}
Suppose that there is available the information that the marginal distributions $F$ and $G$
of $Q$ are \emph{known}. This can be characterized by 
$$
\begin{aligned}
&\int c(x)\,dQ(x,y)=\int c(x)\,dF(x)=0, \quad c \in L_{2,0}(F),\\
&\int d(y)\,dQ(x,y)=\int d(y)\,dG(y)=0,  \quad  d \in L_{2,0}(G). 
\end{aligned}
$$
Bickel, et al.  (1991) and Peng and Schick (2002) constructed   
efficient estimators  of the linear functional $\theta=\int \psi\,dQ$, and proved 
the ASN under the assumption, 
\begin{enumerate}
\item[(K)] There exists $\rho>0$ such that for arbitrary measurable sets $A$ and $B$, 
$$
P(X\in A, Y\in B) \geq \rho F(A)G(B). 
$$
\end{enumerate}
 Bickel, et al.  (1991) showed that the project of $\psi \in L_{2}(Q)$ onto the
sum space $L_{2,0}(F)+L_{2,0}(G)$ uniquely exists. They demonstrated that the asymptotic variance of the
efficient estimator $\tilde\theta$ of $\theta$ can be substantially less than that of the empirical estimator
$\ave \psi(X_j, Y_j)$. For example, they showed that the empirical DF $\ave \1[X_j\leq 1/2, Y_j\leq 1/2]$ of $\theta=P(X\leq 1/2, Y\leq 1/2)$ (taking $\psi_{s,t}(x,y)=\1[x\leq s, y\leq t]$) has three times the asymptotic variance of the efficient estimator $\tilde\theta$ of $\theta$ in the case that $F$ and $G$ 
are uniform distributions over $[0, 1]$ and $X, Y$ are independent 
  
   Here we propose an efficient estimator based on maximum empirical likelihood. 
Employing a basis $\seq{c_k}$ of $L_{2,0}(F)$ and $\seq{d_k}$ of $L_{2,0}(G)$, we can
reduce the uncountably many characterizing equations to countably many ones, 
\bel{km1}
\int c_k(x)\,dF(x)=0, \quad 
\int d_k(y)\,dG(y)=0,  \quad  k=1,2, \dots. 
\ee
Suppose that $F$ and $G$ are continuous. This allows us to take 
$c_k=b_k(F)$ and $d_k=b_k(G)$, where $b_k(t)$ are the trigonometric basis,  
\bel{tb}
b_k(t)=\sqrt{2}\cos(k\pi t), \quad t\in [0, 1], k=1,2, \dots.   
\ee
That is,  $\seq{c_k}$ and $\seq{d_k}$ are bases of $L_{2,0}(F)$
and $L_{2,0}(G)$, respectively.
Using the first $2m_n$ terms as constraints, the EL-weighted estimator of $\btheta$ is
\bel{km}
\hat\theta_n=\Ave \frac{\psi(\Z_j)}{1+\bzeta_n^\top \u_n(\Z_j)},
\end{equation}
where $\u_n(x, y)=(\b_n(F(x))^\top, \b_n(G(y))^\top)^\top$
with $\b_n=(b_1, ..., b_{m_n})^\top$. Using \thmref{3}, we prove  

\Thm{km} Suppose that $F$ and $G$ are continuous. Assume (K).
Then, as $m_n\to\infty$ such that $m_n^4/n\to 0$,
$$
\hat\theta_n=\bar\psi -\bar\varphi_0+o_p(n^{-1/2}),
$$
where $\varphi_0$ is the projection of $\psi$ onto the sum space $L_{2,0}(F)+L_{2,0}(G)$. 
Hence,
$$
\sqrt{n}(\hat\theta_n-\theta)\implies \Normal(0, \Sigma),
$$
where $\Sigma=\Var(\psi(\Z))-\Var(\varphi_0(\Z))$. 

\end{theorem}  

\Rem{eff-km} By Bickel, et al. (1991) (pp. 1328--29), the estimator $\tilde \theta_n$  in \Ref{km} of $\theta=\int\psi(x,y)\,dQ(x,y)$  is semiparametrically
efficient. 

\end{remark}

\medskip
{\sc Proof of \thmref{km}}. We shall rely on \thmref{3}. 
Since $\|\u_n\|\leq 2\sqrt{m_n}$ and $m_n^4/n=o(1)$, 
it follows that \Ref{3a} holds. Thus 
$$
nE(|\bar \W_n-\W_n|_o^2)
\leq E(\|\u_n(\Z)\|^4)
\leq 16m_n^2=o_p(m_n^{-2})
$$
as $m_n^4/n=o(1)$. This shows \Ref{3b}. 
Let 
\bel{km1}
\K_n=\Ave \Big(\psi(\Z_j)\otimes \u_n(\Z_j)- E\big(\psi(\Z_j)\otimes\u_n(\Z_j)\big)\Big).
\ee
It follows from $m_n^2/n=o(1)$ that  \Ref{3c} holds in view of  
$$
E(\|\K_n\|^2)\leq  \frac{1}{n}E(\|\psi(\Z_1)\otimes\u_n(\Z_1)\|^2)
\leq 4\frac{m_n}{n}E(|\psi(\Z_1)|^2)=o(m_n^{-1}).
$$
We are now left to prove the regularity of $\W_n$. 
Since $\b_n$ are the first $m_n$ terms of the orthonormal basis $\seq{b_k}$, 
it follows that $E(\b_n(F(X))\b_n(F(X))^\top)=\I_{m_n}$. 
The same holds for $G$. Let $\C_n=E(\b_n(F(X))\b_n(G(Y))^\top)$. Then
$\W_n$ is the $2m_n \times 2m_n$ dispersion matrix whose (1,1)- and (2, 2)-blocks
are equal to $\bI_{m_n}$ and the (1,2)-block equal to $\C_n$. For $\s, \t \in \cR^{m_n}$ \
with $\|\s\|^2+\|\t\|^2=1$, set $\r=(\s^\top, \t^\top)^\top$.
We have 
\bel{km2}
\r^\top \W_n \r=\|\s\|^2+\|\t\|^2+2\s^\top \C_n\t. 
\ee
By Cauchy inequality,
$$
\begin{aligned}
(\s^\top \C_n \t)^2 
&\leq \s^\top E(\b_n(F(X))\b_n(F(X))^\top)\s\, \t^\top E(\b_n(G(Y))\b_n(G(Y))^\top)\t\\
&=\|\s\|^2\|\t\|^2\leq 1. 
\end{aligned}
$$
It thus follows from \Ref{km2} that $\r^\top \W_n \r\leq 4$ uniformly in $n$ and the 
above $\r$.
 For $a \in L_{2,0}(F)$ and $b \in L_{2,0}(G)$,  (K) implies  
$$
\begin{aligned}
\int(a(x)-b(y))^2\,dQ(x,y) 
&\geq \rho \int (a(x)-b(y))^2dF(x)dG(y)\\
&=\rho\big(\int a^2\,dF+\int b^2\,dG\big).
\end{aligned}
$$
Thus
$$
2\int ab\,dQ \leq (1-\rho)\big(\int a^2\,dF+\int b^2\,dG\big). 
$$
Replacing $a$ with $-a$ yields 
$$
2\int ab\,dQ \geq -(1-\rho)\big(\int a^2\,dF+\int b^2\,dG\big)
$$
 Taking $a=\s^\top\b_n(F)$ and $b=\b_n(G)^\top\t$ and noticing
$$
\int a^2\,dF=\|\s\|^2, \quad \int b^2\,dG=\|\t\|^2, 
$$  
we derive
$$
2\s^\top \C_n\t =2 \int \s^\top\b_n(F(x))\b_n(G(y))^\top\t\,dQ(x,y)
\geq -(1-\rho)(\|\s\|^2 + \|\t\|^2).
$$
By \Ref{km2}, we thus arrive at 
$$
\r^\top \W_n \r\geq \|\s\|^2+\|\t\|^2-(1-\rho)(\|\s\|^2 + \|\t\|^2)
= \rho > 0.
$$
Taken together we prove the regularity of $\W_n$, and apply \thmref{3} to 
complete the proof. 
\qed

\section{Efficient estimation of linear functionals with equal marginals}\label{EQM}
Suppose that the marginal distributions $F$ and $G$ of $X$ and $Y$
are \emph{equal but unknown}.  
This is equivalent to the assertion that 
\bel{eqm1}
E(a_k(X)-a_k(Y))=0, \quad k=1,2, \dots,
\ee
where $\seq{a_k}$ is an orthonormal basis of $L_{2,0}(H)$ with $H=(F+G)/2$. 
Assume that $F$ and $G$ are continuous. This allows us take 
$a_k(x)=b_k(H(x))$ under the assumption $F=G=H$, where
$\seq{b_k}$ is the trigonometric basis in \Ref{tb}. 
As $H$ is unknown, we estimate it by the pooled empirical distribution function, 
$$
\bH(x)=\Ave \half (\1[X_j\leq x]+\1[Y_j\leq x]), \quad x \in \cR.
$$
This gives us computable functions $b_k(\bH(x))$. 
 Let $\u_n(x,y)=\b_n(H(x))-\b_n(H(y)), x,y\in \cR$. This is unknown and can be estimated by $\hat \u_n(x,y)  =\b_n(\bH(x))-\b_n(\bH(y))$.
  Using the first $m_n$ terms as constraints, the EL-weighted estimator 
of $\theta=E(\psi(X, Y))$ is given by
\bel{eqm}
\hat\theta_n=\Ave \frac{\psi(X_j, Y_j)}{1+\hbzeta_n^\top \hat\u_n(X_j, Y_j)},
\end{equation}
where $\hbzeta_n$ is the solution to Eqt \Ref{tele2} with $\u=\hat\u_n$.  

 Peng and Schick (2005) constructed efficient estimators of linear functionals of a bivariate distribution with equal marginals under the condition,  
\bel{lb}
\inf_{a\in A} E[(a(X)-a(Y))^2]>0,
\ee
where $\bA=\{ a\in L_{2,0}(H): \int a^2\,dH=1 \}$ is the unit sphere in $L_{2,0}(H)$. 
They exhibited that the asymptotic variance of an efficient estimator of $\theta$ is about 1/3 of that of the empirical estimator or smaller.

Applying \thmref{4},  we show that $\hat\theta_n$ is efficient. 

\Thm{eqm} Suppose that the distribution functions $F$ and $G$ are equal and continuous.
Assume \Ref{lb}. Then, as $m_n \to \infty$ such that $m_n^6/n\to 0$,
$\hat\theta_n$ given in \Ref{eqm} satisfies 
$$
\hat\theta_n=\bar \psi - \bar \varphi +o_p(n^{-1/2}),
$$
where $\varphi$ is the projection of $\psi$ onto $\bA$. Thus
$$
\sqrt{n}(\hat\theta_n-\theta)\implies \Normal(0, \Sigma),
$$
where  $\Sigma=\Var(\psi)-\Var(\varphi)$.
\end{theorem} 
\def\blam{\bdm{\lambda}} 

\Rem{eff-eqm} By Theorem 3 of Peng and Schick (2005), 
the estimator $\hat\theta_n$ given in \Ref{eqm} of  $\theta=\int\psi(x,y)\,dQ(x,y)$ is semiparametrically efficient. 
\end{remark}

\medskip
{\sc Proof of \thmref{eqm}}. We shall apply \thmref{4}.
Recalling the trigonometric basis $\seq{b_k}$ in \Ref{tb}, one readily verifies that 
$\b_n=(b_1,\ldots,b_{m_n})^\top$ has the properties,   
\bel{trigb}
\|\b_{n}\|\leq ({2m_n})^{1/2}, \quad
\|\b_{n}'\|\leq \sqrt{2}\pi m_n^{3/2}, \quad
\|\b_{n}''\|\leq \sqrt{2}\pi^2 m_n^{5/2},
\ee
where $\b_n'$ and $\b_n''$ denote the first and second order derivatives of $\b$. 

   Recalling $\u_n(x,y)=\b_n(H(x))-\b_n(H(y))$ 
and $\hat\u_n(x,y)=\b_n(\bH(x))-\b_n(\bH(y))$, one gets by the first inequality in \Ref{trigb}  that   
\bel{eqm2}
\|\u_n\|\leq 2\sqrt{2}\sqrt{m_n}, \quad
\|\hat\u_n\|\leq 2\sqrt{2}\sqrt{m_n}. 
\ee
Hence \Ref{4a} holds as $m^4/n=o(1)$. 
Noting $\W_n=E(\u_n(\Z)\u_n(\Z)^\top)$,  one has by \Ref{lb} that 
$$
\blam^\top \W_n\blam=E\big((\blam^\top \b_n(H(X))-\blam^\top \b_n(H(Y)))^2\big)
\geq \inf_{a\in A} E\big(( a(X)- a(Y))^2\big)>0
$$
uniformly in $n$ and $\|\blam\|=1$ as both $\blam^\top \b_n(H(X))$ and $\blam^\top \b_n(H(Y))$ live in $A$.
Moreover,
$$
\blam^\top \W_n\lam \leq 4 E\big((\blam^\top \b_n(H(X)))^2\big)=4.
$$
Thus $\W_n$ is regular. Let 
$$
\hat \W_n=\Ave \hat \u_n(\Z_j)\hat \u_n(\Z_j)^\top, \quad
\bar \W_n=\Ave  \u_n(\Z_j) \u_n(\Z_j)^\top. 
$$
Then by the first equality in \Ref{eqm2},
$$
nE(|\bar \W_n-\W_n|_o^2)
\leq E(|\u_n(\Z_1)|^4)
\leq 64m_n^2.
$$
Hence $\bar \W_n-\W_n=o_p(m_n^{-1})$ as $m_n^4/n=o(1)$. It can be seen
$$
|\bar \W_n-\W_n|_o
\leq D_n+2|\bar \W_n|_o^{1/2}D_n^{1/2},
$$
where $D_n=\ave\|\hat\u_n(\Z_j)-\u_n(\Z_j)\|^2$. 
Thus \Ref{4b} is implied by 
\bel{eqm3}
D_n=o_p(m_n^{-2}). 
\ee 
Using the second inequality in \Ref{trigb}, we derive
$$
\Ave |\b_n(\bH(\Z_j))-\b_n(H(\Z_j))|^2\leq 2\pi^2m_n^3\sup_{t\in\cR}|\bH(t)-H(t)|
=O_p(m^3_n/n).
$$
Hence $D_n=O_p(m^3_n/n)$ and \Ref{eqm3} holds as $m_n^5/n=o(1)$. 
We break 
$$
\Ave \Big(\psi(\Z_j)\otimes \hat \u_n(\Z_j)-E\big(\psi(\Z_j)\otimes \u_n(\Z_j))\Big)
=J_n+K_n, 
$$
where 
$$
\begin{aligned}
& \J_n=\Ave \psi(\Z_j)\otimes \big(\hat \u_n(\Z_j)- \u_n(\Z_j)\big),\\
& \K_n=\Ave \Big(\psi(\Z_j)\otimes \u_n(\Z_j)- E\big(\psi(\Z_j)\otimes\u_n(\Z_j)\big)\Big).
\end{aligned}
$$
By Cauchy inequality, 
$$
\begin{aligned}
 E(\|\J_n\|^2)
&\leq E(|\psi(\Z_1)|^2) \Ave E(\|\hat \u_n(\Z_j)-\u_n(\Z_j)\|^2)
=E(\|\J_n\|^2)E(D_n)\\
&=O(m_n^3/n)=o(m_n^{-1})
\end{aligned}
$$
where the last equality holds as $m_n^4/n=o(1)$. 
We now bound the variance by the second moment and by the first equality in \Ref{eqm2} 
to get
$$
E(\|\K_n\|^2)\leq  \frac{1}{n}E(|\psi(Z_1)\otimes\u_n(Z_1)|^2)
\leq 8\frac{m_n}{n}E(|\psi(Z_1)|^2)=o(m_n^{-1})
$$
as $m_n^2/n=o(1)$. Taken together \Ref{4c} follows. 
We now show \Ref{4e} holds with $\v_n=\u_n$. Using Taylor's expansion, we write
$$
\Ave \big(\b_n(\bH(X_j))-\b_n(H(X_j))\big)
=\L_n+\M_n,
$$
where
$$
\begin{aligned}
&\L_n=\Ave \b_n'(H(X_j))\big(\bH(X_j)-H(X_j)\big),\\
&\M_n=\Ave \b_n''(H_{nj}^*)\big(\bH(X_j)-H(X_j)\big)^2,
\end{aligned}
$$
where $H_{nj}^*$ lies in between $\bH(X_j)$ and $H(X_j)$. 
Using the second inequality in \Ref{trigb}, we get
$$
\begin{aligned}
E\big(\|\L_n\|^2\big)
&\leq \frac{1}{n} E\Big(\|\b_n'(H(X_1))\|^2\big(\bH(X_1)-H(X_1)\big)^2\Big)\\
&\leq 2\pi^2 \frac{m_n^3}{n} \sup_{t \in \cR}\big(\bH(X_1)-H(X_1)\big)^2\\
&=O_p(m_n^3/n^2)=o_p((m_nn)^{-1})
\end{aligned}
$$ 
as $m_n^4/n=o(1)$.  This shows $\L_n=o_p((m_nn)^{-1/2})$. 
Using the third inequality in \Ref{trigb}, one has as $m_n^6/n=o(1)$ that 
$$
\|\M_n\|\leq \sqrt{2}\pi^2 m_n^{5/2}\sup_{t\in\cR}|\bH(t)-H(t)|^2
=O_p(m_n^{5/2}/n)=o_p((m_nn)^{-1/2}).
$$
This yields $\M_n=o_p((m_nn)^{-1/2})$.
Taken together one proves \Ref{4e}. 
This and \Ref{eqm3} imply \Ref{4d} as $m_n^4/n=o(1)$. 
Clearly, $\U_n=\I_{m_n}$ satisfies $|\U_n|_o=1=O(1)$. 
Peng and Schick (2005) showed that the projection of any 
$h \in L_2(Q)$ onto $\bA$ uniquely exists under the assumption \Ref{lb}. Moreover, 
it is clear that $b_k(H(x))-b_k(H(y)), k=1,2, \dots$  
is a basis of $\bA$, so that $[\u_\infty]=\bA$. 
We now apply \thmref{4} to complete the proof. 
\qed 
 \def\M{\mathbf{m}}

\section{Simulations} \label{sim} 
         We ran a simulation study to compare the efficiency of 
the EL-weighted spatial median $\widetilde \M_n$ with the sample spatial median $\M_n$
in the presence of a variety of side information. 
Reported on Tables \ref{ct1}--\ref{lap} are the maximum eigenvalues 
of the asymptotic variance-covariance matrices and their ratios.
Random samples were generated from 2- and 3- dimensional Cauchy distributions,  
Student $t(3)$ with 3 degrees of freedom (df), the copula distributions 
(see the details in the Appendix) and the asymmetric Laplace 
for sample sizes $n=50, 100, 200, 500$. 
Based on  repetitions $M=2000$, we calculated  
the averages of the maximum eigenvalues $\lambda$ and $\tilde\lambda$ (i.e. the spectral
norms) of the asymptotic variance-covariance matrices of $\M_n$ and $\widetilde \M_n$, 
and the ratio $\tilde\lambda/\lambda$.  
A ratio less than one indicates a reduction in the norm of the variance-covariance matrix 
of the EL-weighted spatial median from that of the sample spatial median.

For Table \ref{ct1}, the side information is that the componentwise medians are known. 
For Tables \ref{cau}--\ref{lap},  the information is that one marginal is symmetric about
the origin ($m=1, 3, 5$ constraints considered), for which we looked at both known and unknown
marginal (estimated by the symmetrizied EDF). 

 Observe that for the case of known componentwise medians, 
the efficiency gain of the EL-weighted spatial median over the sample spatial median 
exceeded 80\%; for the case of known or estimated symmetric marginal, 
the efficiency gain is more than 30\%. 
All the ratios considered are substantially smaller than one, indicating  
substantial efficiency gains of the EL-weighted spatial depth over the sample depth.
The simulation results  indicated that the componentwise median 
is less efficient than the spatial median but not that much for the case considered. 

\section{Declaration of interest statement}
\textit{The authors report there are no competing interests to declare.}

\section*{Appendix}\label{app}
{\bf The details of the coupla distributions}.
The 2-dimensional copula distribution has  $N(0, 1)$ and
$t(3)$ marginals with correlation coefficient $0.5$.
The 3-dimensional copula  has  two $N(0, 1)$ marginals with 
correlation coefficient $0.5$ and one $t(3)$ marginal which is correlated with each of $N(0, 1)$ with correlation $0.1$. 
The copula has the joint cumulative distribution function with the 
uniform marginals, where each uniform marginal is defined by applying the
probability integral transform on the cumulative distribution functions  of two $N(0, 1)$ and 
one $t(3)$,  respectively.

We cite Theorem 4 of Wang and Peng (2022) below for  convenience. 
\Thm{4} Suppose $\u_{n}=(u_1, \dots, u_{m_{n}})^\top$ satisfies (C) for each $m=m_n$.    
Let $\hat\u_n$ be an estimator of $\u_n$  such that
\bel{4a}
\maxj \|\hat\u_n(Z_j)\|=o_p(m_n^{-3/2}n^{1/2}), 
\end{equation}
\bel{4b}
|\hat \W_n-\W_n|_o=o_p(m_n^{-1})
\end{equation}
for which the $m_n \times m_n$  dispersion matrices $\W_n$ is regular, 
\bel{4c}
\Ave\pare{\bpsi(Z_j)\otimes \hat\u_n(Z_j)-E\big(\bpsi(Z_j)\otimes \hat\u_n(Z_j)\big)}=o_p(m_n^{-1/2}),
\end{equation}
there exists some measurable function $\v_n$ from $\cZ$ into $\cR^{m_n}$ such that 
(C) is met for every $m=m_n$, the dispersion matrix $\U_n=\W_n^{-1/2}\int \v_n\v_n^\top\,dQ\W_n^{-\top/2}$ satisfies $\U_n=O(1)$, 
\bel{4d}
\Ave E\pare{\|\hat\u_n(Z_j)-\v_n(Z_j)\|^2}=o(m_n^{-1}), \quad \textrm{and} 
\end{equation}
\bel{4e}
\Ave \hat \u_n(Z_j)=\Ave\v_n(Z_j)+ o_p(m_n^{-1/2} n^{-1/2}). 
\ee
Then $\htet$ satisfies, as $m_n$ tends to infinity,  the stochastic expansion, 
\bel{al4}
\htet_n=\bar \bpsi-\bar\bvarphi+o_p(n^{-1/2}),
\end{equation} 
where $\bvarphi=\Pi(\bpsi|[\v_\infty])$ is the projection of $\bpsi$ onto the closed linear span $[\v_\infty]$. 
Thus 
$$
\sqrt{n}(\htet_n-\btheta) \implies \Normal(0, \varSigma).
$$
where  $\varSigma=\Var(\bpsi(Z))-\Var(\bvarphi(Z))$.
\end{theorem} 

\newpage
\begin{table}[!htb]
\centering
\caption{\label{ct1} Simulated maximal eigenvalues $\tilde{\lambda}$ and $\lambda$ 
of the variance-covariance matrices of the EL-weighted and sample spatial medians 
with data generated from a few distributions in the presence of 
known componentwise medians $(0, 0)$ and $(0, 0, 0)$.}
\begin{tabular}{|c|c|c|c|c|c|c|}
\hline
\multicolumn{7}{|c|}{Cauchy}\\
\hline
\multirow{2}{*}{}& \multicolumn{3}{|c|}{dim=2}& \multicolumn{3}{|c|}{dim=3}\\
\hline
& & & & & & \\[-0.9em]
$n$ & $\tilde{\lambda}$&$\lambda$&$\tilde\lambda/\lambda$& $\tilde\lambda$&$\lambda$&$\tilde\lambda/\lambda$\\
\hline
50	&  0.0641	&  0.0094	&  0.1463	&  0.0830	&  0.0111	&  0.1339\\
\hline
100	&  0.0330	&  0.0040	&  0.1209	&  0.0379	&  0.0048	&  0.1260\\
\hline
200	&  0.0157	&  0.0018	&  0.1162	&  0.0207	&  0.0024	&  0.1153\\
\hline
500	&  0.0060	&  0.0007	&  0.1174	&  0.0078	&  0.0009	&  0.1181\\
\hline
\multicolumn{7}{|c|}{Student $t$ (df=3) }\\
\hline
\multirow{2}{*}{}& \multicolumn{3}{|c|}{dim=2}& \multicolumn{3}{|c|}{dim=3}\\
\cline{1-7}
\hline
& & & & & & \\[-0.9em]
$n$ & $\tilde{\lambda}$&$\lambda$&$\tilde\lambda/\lambda$& $\tilde\lambda$&$\lambda$&$\tilde\lambda/\lambda$\\
\hline
50	& 0.0432	& 0.0064	& 0.1477	& 0.0609	& 0.0083	& 0.1363\\
\hline
100	& 0.0218	& 0.0029	& 0.1322	& 0.0281	& 0.0037	& 0.1329\\
\hline
200	& 0.0119	& 0.0014	& 0.1161	& 0.0145	& 0.0017	& 0.1198\\
\hline
500	& 0.0046	& 0.0005	& 0.1096	& 0.0055	& 0.0007	& 0.1257	\\		
\hline
\multicolumn{7}{|c|}{Copula distribution with marginals $N(0,1)$ \& $t(3)$}\\
\hline
\multirow{2}{*}{}& \multicolumn{3}{|c|}{dim=2}& \multicolumn{3}{|c|}{dim=3}\\
\cline{1-7}
\hline
& & & & & & \\[-0.9em]
$n$ & $\tilde{\lambda}$&$\lambda$&$\tilde\lambda/\lambda$& $\tilde\lambda$&$\lambda$&$\tilde\lambda/\lambda$\\
\hline
50	& 0.0523	& 0.0051	& 0.0972	& 0.0542	& 0.0082	& 0.1515\\
\hline
100	& 0.0262	& 0.0021	& 0.0790	& 0.0278	& 0.0036	& 0.1285\\
\hline
200	& 0.0131	& 0.0009	& 0.0715	& 0.0135	& 0.0017	& 0.1291\\
\hline
500	& 0.0054	& 0.0004	& 0.0679	& 0.0055	& 0.0007	& 0.1235	\\				\hline
\multicolumn{7}{|c|}{Asymmetric Laplace}\\
\hline
\multirow{2}{*}{}& \multicolumn{3}{|c|}{dim=2}& \multicolumn{3}{|c|}{dim=3}\\
\cline{1-7}
\hline
& & & & & & \\[-0.9em]
$n$ & $\tilde{\lambda}$&$\lambda$&$\tilde\lambda/\lambda$& $\tilde\lambda$&$\lambda$&$\tilde\lambda/\lambda$\\
\hline
50	& 0.0153	& 0.0021	& 0.1360	& 0.0191	& 0.0024	& 0.1248\\
\hline
100	& 0.0072	& 0.0009	& 0.1213	& 0.0090	& 0.0011	& 0.1209\\
\hline
200	& 0.0035	& 0.0004	& 0.1141	& 0.0043	& 0.0005	& 0.1155\\
\hline
500	& 0.0013	& 0.0001	& 0.1139	& 0.0017	& 0.0002	& 0.1072		\\									
\hline
\end{tabular}
\end{table}

\newpage
\begin{table}[!htb]
\centering
\caption{\label{cau} Same as Table \ref{ct1} except for 
data generated from $3$-dimensional Cauchy 
in the presence of one marginal distribution symmetric about the origin. 
}
\begin{tabular}{|c|c|c|c|c|c|c|c|c|c|}
\hline
\multicolumn{10}{|c|}{One marginal known}\\
\hline
\multirow{2}{*}{}& \multicolumn{3}{|c|}{$m=1$}& \multicolumn{3}{|c|}{$m=3$}& \multicolumn{3}{|c|}{$m=5$}\\
\cline{1-10}
\hline
& & & & & & \\[-0.9em]
$n$ & $\tilde{\lambda}$&$\lambda$&$\tilde\lambda/\lambda$& $\tilde\lambda$&$\lambda$&$\tilde\lambda/\lambda$& $\tilde\lambda$&$\lambda$&$\tilde\lambda/\lambda$\\
\hline
50	& 0.0783	& 0.0499	& 0.6366	& 0.0842	& 0.0527	& 0.6261	& 0.0773	& 0.0525	& 0.6790\\
\hline
100	& 0.0399	& 0.0250	& 0.6277	& 0.0381	& 0.0228	& 0.5989	& 0.0382	& 0.0237	& 0.6213\\\hline
200	& 0.0189	& 0.0119	& 0.6268	& 0.0195	& 0.0116	& 0.5976	& 0.0190	& 0.0116	& 0.6093\\\hline
500	& 0.0074	& 0.0046	& 0.6184	& 0.0082	& 0.0045	& 0.5530	& 0.0075	& 0.0045	& 0.6012\\
\hline
\multicolumn{10}{|c|}{One marginal unknown}\\
\hline
\multirow{2}{*}{}& \multicolumn{3}{|c|}{$m=1$}& \multicolumn{3}{|c|}{$m=3$}& \multicolumn{3}{|c|}{$m=5$}\\
\cline{1-10}
\hline
& & & & & & \\[-0.9em]
$n$ & $\tilde{\lambda}$&$\lambda$&$\tilde\lambda/\lambda$& $\tilde\lambda$&$\lambda$&$\tilde\lambda/\lambda$& $\tilde\lambda$&$\lambda$&$\tilde\lambda/\lambda$\\
\hline
50	& 0.0776	& 0.0518	& 0.6679	& 0.0853	& 0.0533	& 0.6249	& 0.0778	& 0.0532	& 0.6841\\ \hline
100	& 0.0385	& 0.0234	& 0.6082	& 0.0404	& 0.0243	& 0.6016	& 0.0406 & 	0.0241	& 0.5931\\ \hline
200	& 0.0204	& 0.0117	& 0.5720	& 0.0203	& 0.0122	& 0.6020	& 0.0197& 	0.0109	& 0.5538\\ \hline
500	& 0.0073	& 0.0045& 	0.6087	& 0.0079	& 0.0049	& 0.6158	& 0.0078	& 0.0044	& 0.5595\\
\hline
\end{tabular}
\end{table}

\newpage
\begin{table}[!htb]
\centering
\caption{\label{t} 
Same as Table \ref{cau} except for data generated from $t(3)$}
\begin{tabular}{|c|c|c|c|c|c|c|c|c|c|}
\hline
\multicolumn{10}{|c|}{One marginal known}\\
\hline
\multirow{2}{*}{}& \multicolumn{3}{|c|}{$m=1$}& \multicolumn{3}{|c|}{$m=3$}& \multicolumn{3}{|c|}{$m=5$}\\
\cline{1-10}
\hline
& & & & & & \\[-0.9em]
$n$& $\tilde{\lambda}$&$\lambda$&$\tilde\lambda/\lambda$& $\tilde\lambda$&$\lambda$&$\tilde\lambda/\lambda$& $\tilde\lambda$&$\lambda$&$\tilde\lambda/\lambda$\\
\hline
50& 	0.0581& 	0.0384	& 0.6602	& 0.0576	& 0.0357	& 0.6199& 	0.0602& 	0.0385& 	0.6403\\ \hline
100	& 0.0292	& 0.0177& 	0.6069& 	0.0262	& 0.0171	& 0.6519	& 0.0274& 	0.0169	& 0.6178\\ \hline
200 & 	0.0149& 	0.0092& 	0.6204	& 0.0142& 	0.0088	& 0.6209	& 0.0136	& 0.0085	& 0.6239\\ \hline
500	& 0.0055	& 0.0036& 	0.6453	& 0.0057& 	0.0034	& 0.5973	& 0.0056	& 0.0033	& 0.5830\\
\hline
\multicolumn{10}{|c|}{One marginal unknown}\\
\hline
\multirow{2}{*}{}& \multicolumn{3}{|c|}{$m=1$}& \multicolumn{3}{|c|}{$m=3$}& \multicolumn{3}{|c|}{$m=5$}\\
\cline{1-10}
\hline
& & & & & & \\[-0.9em]
$n$& $\tilde{\lambda}$&$\lambda$&$\tilde\lambda/\lambda$& $\tilde\lambda$&$\lambda$&$\tilde\lambda/\lambda$& $\tilde\lambda$&$\lambda$&$\tilde\lambda/\lambda$\\
\hline
50	& 0.0554	& 0.0359	& 0.6481	& 0.0561	& 0.0352	& 0.6278	& 0.0575	& 0.0373	& 0.6477\\ \hline
100	& 0.0291	& 0.0184	& 0.6314	& 0.0288	& 0.0171	& 0.5921	& 0.0286	& 0.0174	& 0.6072\\ \hline
200	& 0.0150	& 0.0096	& 0.6410	& 0.0136& 	0.0087& 	0.6344	& 0.0141	& 0.0086	& 0.6053\\ \hline
500	& 0.0056	& 0.0034	& 0.6112	& 0.0057	& 0.0033	& 0.5732	& 0.0057& 	0.0032	& 0.5622\\
\hline

\end{tabular}
\end{table}

\newpage

\begin{table}[!htb]
\centering
\caption{\label{copula} Same as Table \ref{cau} except 
for data generated from the copula with 
two $N(0,1)$ and one $t(3)$ marginals.}
\begin{tabular}{|c|c|c|c|c|c|c|c|c|c|}
\hline
\multicolumn{10}{|c|}{One marginal known}\\
\hline
\multirow{2}{*}{}& \multicolumn{3}{|c|}{$m=1$}& \multicolumn{3}{|c|}{$m=3$}& \multicolumn{3}{|c|}{$m=5$}\\
\cline{1-10}
\hline
& & & & & & \\[-0.9em]
$n$& $\tilde{\lambda}$&$\lambda$&$\tilde\lambda/\lambda$& $\tilde\lambda$&$\lambda$&$\tilde\lambda/\lambda$& $\tilde\lambda$&$\lambda$&$\tilde\lambda/\lambda$\\
\hline
50	& 0.0545	& 0.0366	& 0.6713	& 0.0541	& 0.0357	& 0.6601	& 0.0527	& 0.0373	& 0.7066\\ \hline
100	& 0.0273	& 0.0181	& 0.6626& 	0.0265	& 0.0183	& 0.6921	& 0.0269	& 0.0174	& 0.6470\\ \hline
200	& 0.0132	& 0.0088& 	0.6665	& 0.0142	& 0.0085	& 0.6005	& 0.0137	& 0.0082	& 0.5995\\ \hline
500	& 0.0054	& 0.0039	& 0.7113	& 0.0053	& 0.0033	& 0.6208	& 0.0053	& 0.0033	& 0.6162\\
\hline
\multicolumn{10}{|c|}{One marginal unknown}\\
\hline
\multirow{2}{*}{}& \multicolumn{3}{|c|}{$m=1$}& \multicolumn{3}{|c|}{$m=3$}& \multicolumn{3}{|c|}{$m=5$}\\
\cline{1-10}
\hline
& & & & & & \\[-0.9em]
$n$& $\tilde{\lambda}$&$\lambda$&$\tilde\lambda/\lambda$& $\tilde\lambda$&$\lambda$&$\tilde\lambda/\lambda$& $\tilde\lambda$&$\lambda$&$\tilde\lambda/\lambda$\\
\hline
50	& 0.0562	& 0.0385& 	0.6841	& 0.0519& 	0.0348& 	0.6698	& 0.0543	& 0.0364	& 0.6707\\ \hline
100	& 0.0275	& 0.0193	& 0.7021	& 0.0272	& 0.0172	& 0.6321	& 0.0267	& 0.0172	& 0.6430\\ \hline
200	& 0.0127	& 0.0089& 	0.7012	& 0.0131	& 0.0085	& 0.6495& 	0.0129	& 0.0082	& 0.6324\\ \hline
500	& 0.0054	& 0.0035& 	0.6427	& 0.0055	& 0.0036& 	0.6451	& 0.0052& 	0.0032	& 0.6187\\
\hline

\end{tabular}
\end{table}

\newpage
\begin{table}[!htb]
\centering
\caption{\label{lap} Same as Table \ref{cau} except 
for data generated from  $3$-dimensional Asymmetric Laplace Distribution.}
\begin{tabular}{|c|c|c|c|c|c|c|c|c|c|}
\hline
\multicolumn{10}{|c|}{One marginal known}\\
\hline
\multirow{2}{*}{}& \multicolumn{3}{|c|}{$m=1$}& \multicolumn{3}{|c|}{$m=3$}& \multicolumn{3}{|c|}{$m=5$}\\
\cline{1-10}
\hline
& & & & & & \\[-0.9em]
$n$& $\tilde{\lambda}$&$\lambda$&$\tilde\lambda/\lambda$& $\tilde\lambda$&$\lambda$&$\tilde\lambda/\lambda$& $\tilde\lambda$&$\lambda$&$\tilde\lambda/\lambda$\\
\hline
50	& 0.0189	& 0.0113	& 0.6006	& 0.0190	& 0.0114	& 0.6004	& 0.0181	& 0.0115	& 0.6335\\ \hline
100	& 0.0090	& 0.0053	& 0.5845	& 0.0093	& 0.0051	& 0.5464	& 0.0089	& 0.0051	& 0.5760\\ \hline
200	& 0.0044	& 0.0026	& 0.5962	& 0.0042	& 0.0023	& 0.5547	& 0.0042	& 0.0023	& 0.5549\\ \hline
500	& 0.0016	& 0.0009& 	0.5867	& 0.0016& 	0.0009	& 0.5627& 	0.0016	& 0.0009	& 0.5525\\
\hline
\multicolumn{10}{|c|}{One marginal unknown}\\
\hline
\multirow{2}{*}{}& \multicolumn{3}{|c|}{$m=1$}& \multicolumn{3}{|c|}{$m=3$}& \multicolumn{3}{|c|}{$m=5$}\\
\cline{1-10}
\hline
& & & & & & \\[-0.9em]
$n$& $\tilde{\lambda}$&$\lambda$&$\tilde\lambda/\lambda$& $\tilde\lambda$&$\lambda$&$\tilde\lambda/\lambda$& $\tilde\lambda$&$\lambda$&$\tilde\lambda/\lambda$\\
\hline
50	& 0.0178	& 0.0118& 	0.6656	& 0.0196	& 0.0118	& 0.6023& 	0.0186& 	0.0130	& 0.6966\\ \hline
100	& 0.0085	& 0.0056& 	0.6565& 	0.0087	& 0.0055& 	0.6346	& 0.0083	& 0.0051	& 0.6162\\ \hline
200	& 0.0042	& 0.0027	& 0.6504	& 0.0043	& 0.0025	& 0.5757	& 0.0044	& 0.0025	& 0.5792\\ \hline
500	& 0.0017	& 0.0011	& 0.6342	& 0.0017	& 0.0010	& 0.5748	& 0.0017	& 0.0010	& 0.5539\\
\hline

\end{tabular}
\end{table}

\end{document}